\documentstyle[12pt]{article}
\newcommand\be{\begin{equation}}
\newcommand\ee{\end{equation}}
\newcommand\bea{\begin{eqnarray}}
\newcommand\eea{\end{eqnarray}}

\newcommand{\acc}{\mbox{\cal{a}}}         

\newtheorem{lemas}{Lemma}
\newenvironment{lema}{\begin{lemas}{\rm\hspace{-2mm}:\hspace{2mm}}}{\end{lemas}}

\newtheorem{teorem}{Theorem}
\newenvironment{teo}{\begin{teorem}{\rm\hspace{-2mm}:\hspace{2mm}}}{\end{teorem}}

\newtheorem{conjetu}{Conjecture}
\newenvironment{conje}{\begin{conjetu}{\rm \hspace{-2mm}:\hspace{2mm}}}{\end{conjetu}}

\newtheorem{corollary}{Corollary}[teorem]
\newenvironment{cor}{\begin{corollary}{\rm \hspace{-2mm}:\hspace{2mm}}}{\end{corollary}}

\begin{document}
\title{Theorems on shear-free perfect fluids with their Newtonian analogues}
\author{Jos\'e M. M. Senovilla and Carlos F. Sopuerta\\
Departament de F\'{\i}sica Fonamental, Universitat de Barcelona \\
Diagonal 647, 08028 Barcelona, Spain \\
and\\
Peter Szekeres \\
Department of Physics and Mathematical Physics, \\
University of Adelaide, South Australia 5000, Australia}
\maketitle

\begin{abstract}
In this paper we provide {\it fully covariant} proofs of some theorems on
shear-free perfect fluids. In particular, we explicitly show that any shear-free
perfect fluid with the acceleration proportional to the vorticity vector 
(including the simpler case of vanishing acceleration) must be either
non-expanding or non-rotating. We also show that these results are not
necessarily true in the Newtonian case, and present an explicit comparison of
shear-free dust in Newtonian and relativistic theories in order to see
where and why the differences appear.
\end{abstract}

PACS: 04.20.-q, 04.20.Jb, 04.40.Nr, 98.80.Hw

\newpage 

\section{Introduction}
This paper deals with shear--free perfect--fluid solutions of Einstein's
field equations.  The motivation for this study comes, on the one hand, from
some studies on kinetic theory (see \cite{TREL} and references therein), and on
the other, from the relations between relativistic cosmology and Newtonian
cosmology. Concerning the former, when we consider isotropic solutions of the 
Boltzmann equation, that is, those for which there is a timelike congruence 
with $\vec{u}$ as the unit tangent vector field such that the distribution 
function has the form $f(x^{a},E)$ with $E \equiv - u^{a}p^{}_{a}$ 
(where $p^{}_{a}$ denotes the particle momentum), two important results 
follow: i) The energy-stress tensor computed from
such a distribution has the perfect-fluid form with respect to
$\vec{u}$ (see for instance \cite{SYNG,TAUW}). ii) The unit tangent vector 
field is shear-free and in addition its expansion $\theta$ and rotation 
$\omega$ satisfy $\omega\theta = 0$ (see the proof in \cite{TREL}).
These results led to the formulation of a conjecture whose origin seems
to be the Ph.\ D. thesis by Treciokas \cite{TREC} (see \cite{LANG} 
for more details). This conjecture can be expressed in the following 
form (here $\varrho$ and $p$ are the energy density and pressure of the
perfect fluid):

\begin{conje} In general relativity, if the velocity vector field
of a barotropic perfect fluid ($\varrho+p \neq 0$ and $p=p(\varrho)$)
is shear-free, then either the expansion or the rotation of the fluid 
vanishes. \label{3trec}
\end{conje} 
While we are still probably a long way from settling the truth or falsity 
of Conjecture 1, it is something short of amazing that such a conjecture 
might be  expected at all in general relativity. Consider for example the 
pressure-free (dust) case for which Ellis \cite{ELL1} showed that $\sigma = 
0 \Longrightarrow  \theta\omega = 0$.  This is a purely local result 
to which no corresponding Newtonian result appears to hold, as 
counterexamples can be explicitly exhibited \cite{HECS}. Ellis's theorem 
holds for arbitrarily weak fields and fluids of arbitrarily low density. 
Why then does the Newtonian approximation fail?

Knowing whether or not this conjecture is true, or at least to what extent
it is valid, might be useful in seeking and studying new perfect-fluid
solutions of Einstein's field equations with a shear-free velocity vector field.
With respect to this subject, there are some interesting studies of shear-%
free perfect-fluid models to be found in \cite{BARN,COL2,COWA,COWH}.
On the other hand, it is important to remark that there are many known
cases which are shear-free {\em and} either rotation-free or expansion-%
free. Some examples are: the Friedmann--Lema\^{\i}tre--Robertson--Walker 
space-times (see for instance \cite{HAE2,ISLA}), the G\"odel solution 
\cite{GOED,HAE2}, spherically symmetric shear-free perfect-fluid solutions 
with expansion and equation of state \cite{KSHM}, Winicour's stationary
dust solutions \cite{WINN}, and other examples of perfect-fluid stationary 
solutions with barotropic equation of state where the velocity vector field 
is aligned with a timelike Killing vector field.

The conjecture has been proved in some special cases.  As far
as we know these cases are the following: 
\begin{enumerate}
\renewcommand{\theenumi}{\roman{enumi}}
\item Spatially homogeneous space-times: Sch\"ucking \cite{SCHU} studied 
the case of dust ($p=0$); Banerji \cite{BANE} studied the case with a 
linear equation of state $p = (\gamma-1)\varrho$ with $\gamma > 1$ and 
$\gamma \neq 10/9$; and finally King and Ellis \cite{KIEL} showed the 
general case with $\varrho+p> 0$.  
\item Ellis \cite{ELL1} proved the conjecture for dust ($p=0$). 
\item Treciokas and Ellis \cite{TREL} showed the case for incoherent 
radiation ($p =1/3\varrho$). 
\item The case in which the acceleration and the vorticity of the
perfect fluid are parallel was shown by White and Collins \cite{WHCO}. 
\item The case in which the magnetic part of the Weyl tensor with respect 
to the velocity vector field vanishes was proved by Collins \cite{COL1}.
\item Carminati \cite{CAR1} showed that for the case of Petrov type N the
conjecture also holds.  
\item Lang and Collins \cite{LANG} proved the same
for the case in which the expansion and the energy density are functionally 
dependent ($\theta=\theta(\varrho)$).
\item Coley \cite{COLE} considered the existence of a conformal Killing
vector field parallel to the velocity vector field $\vec{u}$, proving
also the conjecture for this case. 
\item Finally, there are some recent partial results on Petrov type III by 
Carminati \cite{CAR2} and by Carminati and Cyganowski \cite{CACY}.
\end{enumerate}

It should be stressed, however, that all the above results have been proved
using either a particular tetrad or coordinate system. In this sense, there 
is a clear need\footnote{This was emphasized by G.F.R. Ellis in his plenary
lecture during the past Indian International Conference on Gravitation and
Cosmology (ICGC-95), held in Pune, December 1995. Two of us (JMMS and PS)
were attending this lecture which aroused our interest on the subject
independently.} for a {\it fully covariant} proof of some of the above 
partial theorems. Such covariant proofs are not only desirable on 
aesthetical grounds, but they may also be useful for the deeper 
understanding of why the theorems hold. They may thereby aid in the further 
development of the subject, perhaps helping in eventually proving or 
disproving the above Conjecture 1.

In this paper, we present fully covariant proofs of the theorems in two relevant
cases:  when the acceleration of the fluid vanishes [including the above case
(ii) of dust] and when the acceleration is parallel to the vorticity vector [case
(iv)]. Analogous  covariant proofs can also be given for other cases (see
\cite{SOPU}) by  using similar procedures. In Section 2 we present the main
equations and Sections 3 and 4 are devoted to the proofs of the two 
theorems. Finally Section 5 deals with Newtonian theory, and by attempting
to follow our general relativistic proof, uncovers the reason for the
failure of the Newtonian limit. In addition, all shear-free Newtonian 
universes are revealed by this analysis.

\section{General results on shear-free perfect fluids.}
Let us consider a perfect fluid with unit velocity vector field $\vec{u}$, 
so that the energy-stress tensor reads\footnote{Throughout this paper we
use $a,b,c,\dots,h=1,2,3,4$ for spacetime indices, while we use $i,j,k,\dots=
1,2,3$ for indices in the Newtonian theory, see section 5.}
\be
T^{}_{ab} = \varrho u^{}_{a}u^{}_{b} + p P^{}_{ab}, \label{2temd}
\ee
where $\varrho$ and $p$ are the energy density and the pressure, respectively,
and 
\be
P^{}_{ab} \equiv g^{}_{ab} + u^{}_{a}u^{}_{b}, \hspace{1cm} 
P^{}_{ab} = P^{}_{(ab)}
\label{2proj}
\ee
is the projector orthogonal to $\vec{u}$, which has the standard properties
\be
P^{a}_{\;b} P^{b}_{\;c} = P^{a}_{\;c} , \hspace{3mm}
P^{a}_{\;a} = 3 , \hspace{3mm} P^{a}_{\;b}u^{b} = 0 \, .
\ee

Let us summarize briefly the main concepts for the study of the kinematics of
the velocity $\vec{u}$. The derivative along $\vec{u}$ of any tensor
quantity with components 
$A^{a^{}_{1}\cdots\,a^{}_{p}}_{\;b^{}_{1}\cdots\,b^{}_{q}}$ 
will be denoted by
\be
u^{a}\nabla^{}_{\!a}A^{a^{}_{1}\cdots\,a^{}_{p}}_{\;b^{}_{1}
\cdots\,b^{}_{q}} \equiv \dot{A}^{a^{}_{1}\cdots\,a^{}_{p}}_{\;b^{}_{1}
\cdots\,b^{}_{q}} \; .
\ee
Some different names for this derivative are used in the literature: material
derivative, convective derivative, time derivative with respect to $\vec{u}$,
etc (see \cite{TRUE}). Along this paper we shall always use the term
{\em time propagation} along $\vec{u}$.  
A particular and interesting case is the time propagation along $\vec{u}$
of $\vec{u}$ itself, which is called the {\em acceleration} vector field of
the fluid and is denoted by
\be
\acc^{a} \equiv u^{b}\nabla^{}_{\!b}u^{a} = \dot{u}^{a} .
\ee
>From this definition and taking into account the fact that $\vec{u}$
is a unit vector field, it follows that the acceleration is 
orthogonal to $\vec{u}$ and therefore it is a spacelike
vector field.  Moreover, the integral curves of $\vec{u}$ are geodesic only
when its acceleration vanishes, so that the fluid is said to be {\it geodesic}
if the acceleration vanishes. Physically, the acceleration vector field 
represents the mixed effects of gravitational as well as
inertial forces (see refs.\cite{BEL1,ELL2}).

The spatial part of the covariant derivative of $\vec{u}$ decomposes in general
into its irreducible parts with respect to the rotation group (see 
\cite{ELL2,ELL3,KRSA} and references therein) as follows
\be
\nabla^{}_{b}u^{}_{a}+\acc^{}_{a}u^{}_{b} =
\frac{1}{3}\theta P^{}_{ab} + \sigma^{}_{ab}+\omega^{}_{ab} . \label{2cod1}
\ee
where
\be
\theta = \nabla^{}_{\!a}u^{a} ,   \label{2exra} 
\ee
\be
\sigma^{}_{ab} = P_{(a}^{\;c}P_{b)}^{\;d}\nabla^{}_{\!d}u^{}_{c}
-\frac{1}{3}\theta P^{}_{ab} ,  \hspace{5mm} \sigma^{}_{ab} =
\sigma^{}_{(ab)} ,\hspace{3mm} \sigma^{a}_{\;a} = 0 , 
\hspace{3mm} \sigma^{}_{ab}u^b = 0 , \label{2shra} 
\ee
\be
\omega^{}_{ab} = P_{[a}^{\;c}P_{b]}^{\;d}\nabla^{}_{\!d}u^{}_{c} ,
\hspace{5mm} \omega^{}_{ab} = \omega^{}_{[ab]} ,
\hspace{3mm} \omega^{}_{ab}u^b = 0 . \label{2wwab}
\ee
These objects together with the acceleration $\vec{\acc}$ form 
the kinematical quantities of $\vec{u}$.
The commonly used names for them are: {\em expansion rate} 
for $\theta$, {\em shear rate} for $\sigma^{}_{ab}$ and {\em rotation
rate} for $\omega^{}_{ab}$.

We can also introduce the {\em vorticity} vector field by
\be 
\omega^{a} \equiv \frac{1}{2}\eta^{abcd}u^{}_{b}\omega^{}_{cd} 
\label{2vort}
\ee
from where we see that the vorticity is orthogonal to $\vec{u}$, that is,
$\omega_a u^a =0$. Moreover, we can invert (\ref{2vort}) in order to obtain the
rotation rate in terms of the vorticity
\be
\omega^{}_{ab} = \eta^{}_{abcd}\omega^{c} u^{d}  \label{2rovo}
\ee
where $\eta^{}_{abcd}$ is the canonical volume element in the spacetime (see,
for instance, \cite{EISE,HAE2,KSHM}).
>From (\ref{2rovo}) it follows directly that $\omega^{}_{ab}\omega^{b} = 0$.
We also define the rotation scalar by
\be
\omega^{2} \equiv \frac{1}{2}\omega^{ab}\omega^{}_{ab} = \omega^{a}
\omega^{}_{a} ,
\ee
and, given that $\omega_{ab}$ is a spatial tensor (that is, completely
orthogonal to $\vec{u}$), $\omega^{2}$ is non--negative and the following
equivalences hold:
\be 
\omega = 0 \hspace{5mm} \Longleftrightarrow \hspace{5mm} \omega^{a} = 0 
\hspace{5mm} \Longleftrightarrow \hspace{5mm} \omega^{}_{ab} = 0 .
\ee
>From (\ref{2rovo}) we can immediately get the following standard identities
\be
\omega_a^{\;c}\,\omega_c^{\;b}=\omega_a\omega^b-\omega^2 P_a^{\;b} \hspace{5mm}
\Longrightarrow \hspace{5mm}\omega_a^{\;c}\,\omega_c^{\;d}\,\omega_d^{\;b}=
-\omega^2\omega_a^{\;b} \, . \label{id}
\ee
Finally, let us remark an important fact concerning $\omega_{ab}$. 
Taking into account the following expression which comes directly from
(\ref{2cod1})
\be
\nabla^{}_{[a}u^{}_{b]}=\acc^{}_{[a}u^{}_{b]}+\omega^{}_{ba} ,
\label{2du1f}
\ee
it follows that 
\be
u^{}_{[c}\nabla^{}_{a}u^{}_{b]}=0 \hspace{5mm}  \Longleftrightarrow
\hspace{5mm} \omega^{}_{ab}=0 , \label{frob}
\ee
and using Frobenius's theorem (see \cite{FLAN,KSHM}), this means that there 
exist locally functions $f$ and $h$ such that $u_a=h\nabla_a f$, (or 
equivalently, $\vec{u}$ generates orthogonal spacelike hypersurfaces), if and
only if the rotation vanishes. 

The fluid (or also $\vec{u}$) is said to be {\it shear-free} if 
\be
\sigma_{ab}=0
\ee
which we assume from now on. In this case, equation (\ref{2cod1}) becomes
simply
\be
\nabla^{}_{b}u^{}_{a} =\frac{1}{3}\theta P^{}_{ab} +\omega^{}_{ab}
- \acc^{}_{a}u^{}_{b} . \label{2cod2}
\ee
Then, by using the Ricci identities for $\vec{u}$
\be
\left(\nabla^{}_{\!c}\nabla_{\!d}^{} - \nabla^{}_{\!d}\nabla^{}_{\!c} 
\right) u^{a} = R^{a}_{\;bcd} u^{b}  \label{2ridu}
\ee
we can compute the time propagation along $\vec{u}$ of (\ref{2cod2}),
which produces the following set of {\it evolution equations}
\cite{EHL1,EHL3,HAWK} (we use units with $8\pi G=c=1$)
\be
\dot{\theta} + \frac{1}{3}\theta^{2} - 2 \omega^{2} - \nabla^{}_{\!a} \acc^{a}
+  \frac{1}{2} \left(\varrho + 3 p \right) = 0 ,  \label{2rayc} 
\ee
\be
P^{\;c}_{a}P^{\;d}_{b}\dot{\omega}^{}_{cd}+\frac{2}{3}\theta \omega^{}_{ab}-
P^{\;c}_{[a}P^{\;d}_{b]}\nabla^{}_{\!d}\acc^{}_{c} = 0 , \label{2ropo} 
\ee
\be
E^{}_{ab} =-\omega^{}_{a}\omega^{}_{b}+\frac{1}{3}\omega^{2}P^{}_{ab}+
\acc^{}_{a}\acc^{}_{b} + P^{\;c}_{(a}P^{\;d}_{b)} 
\nabla^{}_{\!c}\acc^{}_{d}-\frac{1}{3}P^{}_{ab}\nabla^{}_{\!c}\acc^{c} . 
\label{3elwe} 
\ee
Here $E^{}_{ab}$ is the so-called {\em electric} part of the Weyl tensor with
respect to $\vec{u}$. This and the {\em magnetic} part $H^{}_{ab}$
relative to $\vec{u}$ determine completely the Weyl tensor $C^{}_{acbd}$,
and are defined respectively by
\be
E^{}_{ab} \equiv C^{}_{acbd}u^{c}u^{d} \hspace{5mm} \Longrightarrow 
\hspace{5mm}
E^{}_{ab} = E^{}_{(ab)}, \hspace{2mm} E^{}_{ab}u^{b} = 0, 
\hspace{2mm} E^{a}_{\;a} = 0 , \label{2elec}
\ee
\be
H^{}_{ab} \equiv \frac{1}{2}\eta^{\;\;\;cd}_{ae}C^{}_{cdbf}u^{e}u^{f} 
\hspace{5mm} \Longrightarrow \hspace{5mm} H^{}_{ab} = H^{}_{(ab)}, \hspace{2mm}
H^{}_{ab}u^{b} = 0, \hspace{2mm} H^{a}_{\;a} = 0 . \label{2magn}
\ee
These tensors were first introduced by Matte \cite{MATT} in the
context of the study of gravitational radiation.

Equation (\ref{2rayc}) is the famous Raychaudhuri equation \cite{RAYC}, and
Eq.(\ref{2ropo}) can also be given in terms of the vorticity as follows
\be
P^{a}_{\;b}\dot{\omega}^{b} + \frac{2}{3}\theta\omega^{a} - 
\frac{1}{2}\eta^{abcd}u^{}_{b}\nabla^{}_{\!d}\acc^{}_{c} = 0 . \label{2vopo}
\ee
Eq.(\ref{3elwe}) comes directly from the time-propagation of the
absence of shear. The whole set (\ref{2rayc},\ref{2ropo},\ref{3elwe})
contains 9 of the 18 independent components of the Ricci identities
(\ref{2ridu}). The remaining 9 components are usually called {\it constraint
equations} and are given by \cite{EHL1,EHL3,HAWK}
\be
2 P^{ab}\nabla^{}_{\!b} \theta +3 P^{a}_{\;b}\nabla^{}_{\!d}
\omega^{bd} +3 \omega^{a}_{\;b}\acc^{b} = 0 , \label{3main}
\ee
\be
\nabla^{}_{\!a} \omega^{a} = 2 \acc^{}_{a} \omega^{a} \hspace{5mm}
\Longleftrightarrow \hspace{5mm} \eta^{abcd}u^{}_{a}\left(\nabla^{}_{\!b}
\omega^{}_{cd}-\acc^{}_{b}\omega^{}_{cd}\right) = 0, \label{2divv} 
\ee
\be
H^{}_{ab} = 2\omega^{}_{(a}\acc^{}_{b)}-(\acc^{c}\omega^{}_{c})P^{}_{ab}+
P^{\;c}_{(a}P^{\;d}_{b)}\nabla^{}_{\!c}\omega^{}_{d} , \label{3mawe}
\ee
where in order to write down the magnetic part $H^{}_{ab}$ in the form
(\ref{3mawe}) we have used the identity
\be
\omega^{}_{ab}\nabla^{}_{\!c}\omega^{bc} = -\omega^{2}\acc^{}_{a} +
(\acc^{b}\omega^{}_{b})\omega^{}_{a} + P^{\;b}_{a}
\omega^{c}\nabla^{}_{\!c}\omega^{}_{b} - P^{\;b}_{a}\omega^{c}
\nabla^{}_{\!b}\omega^{}_{c} . \label{3uti1}
\ee

Finally, the conservation equations $\nabla^{}_{\!b}T^{ab} = 0$ for the
energy-stress tensor (\ref{2temd}) read 
\be
\dot{\varrho} + \left(\varrho+p\right) \theta =0, \label{2ropu}
\ee
\be
\left(\varrho+p\right)\acc^{a}+P^{ab}\nabla_{b}p = 0 . \label{2moco}
\ee 
Now, we are ready to present the fully covariant proofs of the main theorems.

\section{Case without acceleration: $\vec{\acc} = \vec{0}$.}
In this section we present the coordinate-- and tetrad--free proof
of Conjecture \ref{3trec} for the case of vanishing acceleration
\be
\vec{\acc}=\vec{0}, \label{new}
\ee
which includes the particular case of dust because of (\ref{2moco}). First of
all, we present some lemmas which will help us in the proof of the theorem.

\begin{lema} If there exists a function $f$ satisfying $P^{ab}\nabla^{}_b f=0$
then either $f=$const.\ or the rotation vanishes.
\end{lema}

\noindent
{\it Proof:} \hspace{3mm}
There are two possibilities, either $\dot{f}$ vanishes or not. If it does, 
then $f=$const.\ clearly. If it does not, then we can write
\be 
u^{}_a = -\frac{1}{\dot{f}} \nabla^{}_a f .\label{ugrad}
\ee
But then, as we have seen in the comments following equation (\ref{frob}),
this implies $\omega^{}_{ab}=0$, proving the lemma.

\begin{lema} If the perfect fluid is geodesic then either the pressure $p$ 
is constant or the rotation vanishes. 
\end{lema}

\noindent
{\it Proof:} \hspace{3mm} 
The proof of this lemma comes from the energy-stress conservation
equation (\ref{2moco}), which by using (\ref{new}) becomes
$P^{ab}\nabla^{}_{\!b}p = 0$. Then Lemma 1 implies the result.

\begin{lema} If the perfect fluid is geodesic and shear-free, and there 
exist constants $c_1$ and $c_2$ with $c_2\neq 0$ such that
 \be
\varrho =(c_1 -1)\, p+c_2 \, \omega^2 \, , \label{l3}
\ee
then either the rotation or the expansion vanishes.
\end{lema}

\noindent
{\it Proof:} \hspace{3mm} First of all, the geodesic condition implies that
the time-propagation equations (\ref{2ropo}) and (\ref{2vopo}) for the 
rotation and the vorticity reduce to
\bea
P^{\;c}_{a}P^{\;d}_{b}\dot{\omega}^{}_{cd}=\dot{\omega}^{}_{ab}=
-\frac{2}{3}\theta \omega^{}_{ab} , \hspace{1cm} \label{wwdot} \\
P^{a}_{\;b}\dot{\omega}^{b}=\dot{\omega}^{a} =- \frac{2}{3}\theta\omega^{a}
\hspace{5mm} \Longrightarrow \hspace{5mm}
\dot{\omega}=-\frac{2}{3}\theta \omega . \label{wdot}
\eea
>From Lemma 2 either the rotation vanishes or the pressure
is constant. Thus, we need only consider the case $p=$const.
The time propagation of (\ref{l3}) then gives, on using (\ref{2ropu}), 
and (\ref{wdot}),
\bea
\theta \left(c_1 p - \frac{c_2}{3}\omega^2\right) =0 .
\eea
If $\theta =0$ we are done. If the term in brackets vanishes, then 
$\omega$ is constant, and (\ref{wdot}) gives $\theta \, \omega =0$ as 
required.
\vspace{5mm}
\noindent We now pass to the proof of the theorem.
\begin{teo} Every shear-free and geodesic perfect fluid must be either
expansion-free or rotation-free, i.e.
\begin{eqnarray*}
\sigma^{}_{ab}=0, \hspace{2mm} \acc^b=0 \hspace{5mm} \Longrightarrow
\hspace{5mm} \omega \theta =0 .
\end{eqnarray*}
\label{3elli}
\end{teo}

\noindent This theorem was first shown by Ellis \cite{ELL1} for the dust 
case ($p = 0$) and it was completed for the case of constant pressure by
White and Collins \cite{WHCO}. Both theorems were proved using 
the so-called tetrad formalism (see, for instance, \cite{MACC}).\\[3mm]
{\it Proof:} \hspace{3mm} From Lemma 2 it is clearly only necessary to
consider the case $p$=const. Moreover, from (\ref{2cod2}) 
restricted to the case with (\ref{new}) we get immediately, for any 
function $f$, the following identity
\be
\left(P^{ab}\nabla^{}_b f\right)\dot{}=P^{ab}\nabla^{}_b \dot{f}+
\omega^{ab}\nabla^{}_b f -\frac{\theta}{3}P^{ab}\nabla^{}_b f \, .\label{dotid}
\ee
Using this identity together with the relations (\ref{wwdot}-\ref{wdot}), 
the time propagation of Equation (\ref{3main}) restricted to
the case $\acc^{b}=0$ results in
\be
2 P^{\;b}_{a}\nabla^{}_{\!b}\varrho - 13 P^{\;b}_{a}\omega^{c}
\nabla^{}_{\!b}\omega^{}_{c}-3 P^{\;b}_{a}\omega^{c}\nabla^{}_{\!c}
\omega^{}_{b} = 0.  \label{3tpmp}
\ee
On using (\ref{3main}) contracted with $\omega^{}_{ac}$ and the
identity (\ref{3uti1}), this equation can be rewritten as
\be
P^{\;b}_{a}\nabla^{}_{\!b}\varrho -8\omega P^{\;b}_{a}\nabla^{}_{\!b}\omega
+\omega^{\;b}_{a}\nabla^{}_{\!b}\theta =0 . \label{key}
\ee
This key equation provides an {\it algebraic} relation between the spatial
gradients of density, rotation and expansion. By contracting it with
$\omega^{ca}$ we also get
\be
\omega^{\;b}_{a}\nabla^{}_{\!b}\varrho -
8\omega \, \omega^{\;b}_{a}\nabla^{}_{\!b}\omega
+\omega^{\;c}_{a}\omega^{\;b}_{c}\nabla^{}_{\!b}\theta =0 . \label{key2}
\ee
The next step is the computation of the time propagation of Equation
(\ref{key}). Making use of (\ref{wwdot}-\ref{wdot}), (\ref{dotid}),
(\ref{2ropu}), (\ref{key2}) and the Raychaudhuri equation (\ref{2rayc}) 
this leads to
\be
\left(\varrho +p - \frac{16}{3}\omega^2\right)P^{\;b}_{a}\nabla^{}_{\!b}\theta=
\frac{1}{2}\omega^{\;c}_{a}\omega^{\;b}_{c}\nabla^{}_{\!b}\theta +
\frac{\theta}{3}P^{\;b}_{a}\nabla^{}_{\!b}\varrho \, \, . \label{31}
\ee
Time-propagating again this relation and taking into account the key equations
(\ref{key}) and (\ref{key2}) together with
(\ref{wwdot}-\ref{wdot}), (\ref{dotid}), (\ref{2rayc}), and (\ref{31}) itself
we arrive at
\bea
\frac{1}{2}\left(\varrho +p - \frac{16}{3}\omega^2\right)
\omega^{\;b}_{a}\nabla^{}_{\!b}\theta + \theta \left[\frac{112}{9}\omega^2-
\frac{5}{3}\left(\varrho+p\right)\right]P^{\;b}_{a}\nabla^{}_{\!b}\theta +
\hspace{30mm} \nonumber \\
+\left(\frac{5}{9}\theta^2-\frac{2}{3}\omega^2+\frac{\varrho+3p}{6}\right)
P^{\;b}_{a}\nabla^{}_{\!b}\varrho+\frac{7}{6}\theta
\omega^{\;c}_{a}\omega^{\;b}_{c}\nabla^{}_{\!b}\theta -\frac{1}{4}
\omega^{\;d}_{a}\omega^{\;c}_{d}\omega^{\;b}_{c}\nabla^{}_{\!b}\theta =0 \, .
\hspace{5mm}
\eea
Multiplying this by $\theta$ in order to use (\ref{31}) again, and 
re--ordering terms it follows that
\bea
\frac{\theta}{2}\left(\varrho +p - \frac{16}{3}\omega^2\right)
\omega^{\;b}_{a}\nabla^{}_{\!b}\theta -\frac{\theta}{4}
\omega^{\;d}_{a}\omega^{\;c}_{d}\omega^{\;b}_{c}\nabla^{}_{\!b}\theta +
\left(\frac{\theta^2}{3}+\omega^2-\frac{\varrho+3p}{4}\right)
\omega^{\;c}_{a}\omega^{\;b}_{c}\nabla^{}_{\!b}\theta + \nonumber \\
+\left[\frac{32}{9}\theta^2\omega^2+ \left(\varrho + p -\frac{16}{3}\omega^2
\right)\left(\frac{\varrho+3p}{2}-2\omega^2\right)\right]P^{\;b}_{a}
\nabla^{}_{\!b}\theta =0 \, . \label{F}
\eea
Contracting with $\omega^a$ we find the simple relation
\be
\left[\frac{32}{9}\theta^2\omega^2+ \left(\varrho + p -\frac{16}{3}\omega^2
\right)\left(\frac{\varrho+3p}{2}-2\omega^2\right)\right]
\omega^b\nabla^{}_b \theta = 0,  \label{F2}
\ee
so that two different possibilities appear:

1) If $\omega^b\nabla^{}_b \theta \neq 0$, then 
\be
\frac{32}{9}\theta^2\omega^2+ \left(\varrho + p -\frac{16}{3}\omega^2
\right)\left(\frac{\varrho+3p}{2}-2\omega^2\right) = 0 \, . \label{F4}
\ee
The time propagation of this equation leads on using Eqns. (\ref{2rayc}),
(\ref{2ropu}), (\ref{wdot}) and (\ref{F4}) itself, either to $\theta =0$ 
(in which case obviously $\theta\omega =0$) or
\bea
\frac{64}{9}\omega^4 -2 \varrho \, \omega^2-\frac{38}{3}p \, \omega^2 +
p(\varrho +p) =0 .\label{cond1a}
\eea
Once more time-propagating this relation we finally arrive at
\bea
\left(\varrho +\frac{7}{3}p-\frac{40}{9}\omega^2\right)
\, \omega^2 =0 \, .\label{cond1b}
\eea
Thus, either $\omega =0$ and we are done, or the term in brackets vanishes
in which case $\omega \theta =0$ by Lemma 3.

2) If $\omega^b\nabla^{}_b \theta =0$, then Equation 
(\ref{F}) together with the identities (\ref{id}) lead to
\bea
\frac{\theta}{2}\left(29\omega^2-6(\varrho+p)\right)
\omega^{\;b}_{a}\nabla^{}_{\!b}\theta = \hspace{8cm} \nonumber \\
=\left[\frac{58}{3}\theta^2\omega^2 +\left(2\omega^2 -\frac{\varrho+3p}{2}
\right)\left(29\omega^2-6(\varrho+p)\right)\right]
P^{\;b}_{a}\nabla^{}_{\!b}\theta \, .\hspace{5mm}  \label{F3}
\eea
But the two vectors $P^{\;b}_{a}\nabla^{}_{\!b}\theta$ and
$\omega^{\;b}_{a}\nabla^{}_{\!b}\theta$ are orthogonal to each other, as
follows immediately from (\ref{2rovo}),
\bea
\omega^{\;b}_{a}\nabla^{}_{\!b}\theta =\omega_{ac}P^{cb}\nabla^{}_b\theta =
\eta^{}_{acde}\omega^d u^e P^{cb}\nabla^{}_b\theta \, .
\eea
Hence the left-hand side of relation (\ref{F3}) must vanish, and there are
three possibilities.  Either (a) $\theta=0$, in which case the theorem is 
proved, or (b) the term in brackets vanishes whence the theorem follows 
from Lemma 3, or (c) $\omega^{\;b}_{a}\nabla^{}_{\!b}\theta =0$. In this
case using the condition defining case (2), namely $\omega^b\nabla^{}_b 
\theta =0$, it follows that either $P^{\;b}_{a}\nabla^{}_{\!b}\theta =0$ or 
$\omega^b = 0$.  If the latter holds then $\omega=0$ and the theorem is 
proved, while the former implies through (\ref{31}) and (\ref{key}) that 
$P^{\;b}_{a}\nabla^{}_{\!b}\varrho=P^{\;b}_{a}\nabla^{}_{\!b}\omega =0$.
Then setting $f=\omega$ in Lemma 1 gives $\omega$ is constant and 
Equation (\ref{wdot}) implies that $\theta \omega =0$. Hence the 
theorem holds in this case too.\\[2mm]
This finishes the proof of Theorem \ref{3elli}.

It must be remarked that Theorem 1 does not need the assumptions of a
ba\-ro\-tro\-pic equation of state and $\varrho +p\neq 0$. In particular, 
this theorem also holds for Einstein's spaces (including vacuum) in which 
there is a geodesic shear-free unit timelike vector field.

\begin{cor} If $u^a$ is a shear-free geodesic vector field in any Einstein 
space ($G_{ab}=\lambda g_{ab}$), then $\theta\omega = 0$. \end{cor}

\noindent {\it Proof: }\hspace{3mm} Essentially this is the case
\[\varrho = -p = -\lambda = \mbox{const}.\]
Equation (\ref{F2}) is replaced by the simpler
\be \omega^2\left[\frac{2}{3}\theta^2 +\rho +2\omega^2\right] \omega^b \nabla_b\theta = 0 \label{F20}. \ee
If the expression in brackets vanishes we have by essentially the same
discussion as given in case (1) above,
\[\varrho + 2\omega^2 = 0\]
whence $\theta\omega=0$ follows by Lemma 3.
On the other hand if $\omega^b\nabla_b\theta = 0$ then the proof follows
exactly as for case (2), with appropriate simplifications for example
in Equation (\ref{F3}).

\section{Case in which the acceleration and the vorticity vector
fields are parallel: $\vec{\acc}\; \|\; \vec{\omega}$.}

In this section we study the case when the acceleration vector field
$\vec{\acc}$ and the vorticity vector field $\vec{\omega}$ are parallel, that
is to say, when 
\be
\vec{\acc} = \psi \vec{\omega}, \hspace{5mm} \psi \neq 0 . \label{3fass}
\ee
Taking into account Theorem 1 we will consider $\psi$ as an arbitrary 
{\it non-vanishing} function. This also means by (\ref{2moco}) that, 
whenever there is a barotropic equation of state $p=p(\varrho)$, we have
\be
p' \neq 0 , \hspace{1cm} \mbox{where} \hspace{1cm} 
p' \equiv dp(\varrho)/d\varrho .
\ee
As in the previous section, we collect some useful lemmas before we prove the
theorem.

\begin{lema} For every shear-free perfect fluid with a barotropic equation of
state and $\varrho +p\neq 0$, we have
\be
P^{\;c}_{[b}P^{\;d}_{a]}\nabla^{}_{\!c}\acc^{}_{d} = p'\theta\omega^{}_{ab} ,
\label{3anti}
\ee
\be
P^{a}_{\;b}\dot{\acc}^{b} = \left[p'-\frac{1}{3}-(\varrho+p)\frac{p''}{p'}
\right]\theta\acc^{a}+\omega^{a}_{\;b}\acc^{b}+ p'P^{ab}\nabla^{}_{\!b}\theta . \label{3tpac}
\ee
\end{lema}

\noindent
{\it Proof:} \hspace{3mm} If we introduce the equation of state $p=p(\varrho)$
into relation (\ref{2moco}) and use (\ref{2ropu}) we obtain
\be
\acc^{}_a = p' \, \theta u^{}_a -(\varrho+p)^{-1}\nabla^{}_a p
=p'\left(\theta u^{}_a -(\varrho+p)^{-1}\nabla^{}_a \varrho\right) \, 
,\label{a}
\ee
from which it is easily derived that
\be
P^{\;c}_{[b}P^{\;d}_{a]}\nabla^{}_{\!c}\acc^{}_{d}=p' \,  \theta \,
P^{\;c}_{[b}P^{\;d}_{a]}\nabla^{}_{\!c}u^{}_{d} \, ,
\ee
and now using (\ref{2du1f}) we arrive at (\ref{3anti}). On the other hand,
the time-propagation of (\ref{a}), on using (\ref{3main}) and (\ref{2moco}),
leads directly to (\ref{3tpac}).

\begin{lema} For every shear-free perfect fluid with a barotropic equation of
state and $\varrho +p\neq 0$ , the time propagation of the rotation and
vorticity are
\be
P^{\;c}_{a}P^{\;d}_{b}\dot{\omega}^{}_{cd}=
\left(p'-\frac{2}{3}\right)\theta \omega^{}_{ab} ,
\ee
\be
P^{a}_{\;b}\dot{\omega}^{b} = \left(p'-\frac{2}{3}\right)\theta\omega^{a}
\hspace{5mm} \Longrightarrow \hspace{5mm} \dot{\omega} =  \left(p'-
\frac{2}{3}\right)\theta\omega \, .\label{3tpvo}
\ee
\end{lema}

\noindent
{\it Proof:} \hspace{3mm} The proof is straightforward on introducing 
(\ref{3anti}) into equations (\ref{2ropo}) and (\ref{2vopo}), respectively.

\begin{teo} Every shear-free perfect fluid with a barotropic equation of
state and $\varrho +p\neq 0$ in which the acceleration and vorticity vector
fields are parallel is either expansion-free or
rotation-free. \label{3whco}
\end{teo}

\noindent The proof of this theorem was given by White and Collins 
\cite{WHCO} by using the tetrad formalism \cite{MACC}. Here, we present 
a very simple covariant proof of this result.\\[2mm]
{\it Proof:} \hspace{3mm}
>From the expression (\ref{2rovo}) of the rotation rate $\omega^{}_{ab}$ in
terms of $\vec{\omega}$, taking into account (\ref{3fass}), and using equation
(\ref{3anti}) of lemma 4 we can write
\be
P^{}_{ab}\nabla^{}_{\!d}\omega^{bd} = \eta^{}_{abcd}u^{b}\nabla^{c}\left(
\psi^{-1}\acc^{d}\right) = -\psi^{-1}\left(\omega^{\;b}_{a}\nabla^{}_{\!b}
\psi+2p'\theta\omega^{}_{a}\right) , \label{c}
\ee
which allows us to express Equation (\ref{3main}) in the form
\be
P^{ab}\nabla^{}_{\!b}\theta = 3\psi^{-1}\left(\frac{1}{2}\omega^{ab}
\nabla^{}_{\!b}\psi + p'\theta\omega^{a}\right) . \label{3exg1}
\ee
Introducing this into expression (\ref{3tpac}) of Lemma 4 and using
$\omega^{a}_{\;b}\acc^{b}=\psi \, \omega^{a}_{\;b}\omega^{b}=0$ we obtain 
\be
P^{a}_{\;b}\dot{\acc}^{b} = \left[p'-\frac{1}{3}-(\varrho+p)\frac{p''}{p'}
+3\left(\frac{p'}{\psi}\right)^{2}\right]\theta\acc^{a} +
\frac{3}{2}\psi^{-1}p'\omega^{ab}\nabla^{}_{\!b}\psi \, . \label{3tpa1}
\ee
However another expression for the time propagation of the acceleration 
vector field $\vec{\acc}$ can be obtained by putting the
assumption (\ref{3fass}) into Equation (\ref{3tpvo}) of lemma 5.
The result is 
\be 
P^{a}_{\;b}\,\dot{\acc}^{b} = \left[\left(p'-\frac{2}{3}\right)\theta+
\frac{\dot{\psi}}{\psi}\right] \acc^{a} . \label{3tpa2}
\ee
Comparison of Equations (\ref{3tpa1}) and (\ref{3tpa2}) leads to
\bea
\omega^{ab}\nabla^{}_{\!b}\psi = 0 , \hspace{2cm} \label{b} \\
\dot{\psi} = \left[3\left(\frac{p'}{\psi}\right)^{2}+\frac{1}{3}-
(\varrho+p)\frac{p''}{p'}\right]\theta\psi \, , \label{3psid}
\eea
and the first of these, through Equation (\ref{3exg1}), leads in turn to
the following simple formula
\be
\nabla^{}_a \theta = -\dot{\theta} u^{}_a + 3\psi^{-1}p'\theta \omega^{}_a .
\label{3dthe}
\ee
Another consequence of (\ref{b}) is that relation (\ref{c}) simplifies to
\be
P^{}_{ab}\nabla^{}_{\!d}\omega^{bd} = -2\frac{p'}{\psi}\theta\omega^{}_{a} , 
\label{seke}
\ee
and on contracting with $\omega^{\;a}_{c}$ and using the identity 
(\ref{3uti1}) we get
\be 
P^{ab}\omega^{c}\nabla^{}_{\![c}\omega^{}_{b]} = 0 . \label{notie}
\ee
Considering this expression together with the
magnetic part of the Weyl tensor (\ref{3mawe}) and the Ricci identities
for $\omega^{ab}$, the time propagation of (\ref{seke}) gives
\be
\dot{\theta} = 3\left(\frac{p'}{\psi}\right)^{2}\theta^{2} .
\label{3dott}
\ee
Putting this into equation (\ref{3dthe}) and applying the
integrability conditions $\nabla^{}_{\![a}\nabla^{}_{\!b]}\theta = 0$
for $\theta$, we obtain that either $\theta =0$, which proves the theorem, or
\be
\frac{p'}{\psi^{2}}\omega^{a}\nabla^{}_{\!a}\psi = \left[ 3\left(
\frac{p'}{\psi}\right)^{2}+\frac{1}{3}-(\varrho+p)\frac{p''}{p'}\right]
\omega^{2} \, .
\ee
On the other hand, introducing (\ref{3dott}) into the Raychaudhuri 
equation (\ref{2rayc}) and using the
assumptions of this case we obtain another expression for 
$\omega^{a}\nabla^{}_{\!a}\psi$, 
\be
\omega^a\nabla^{}_a \psi =\left[3\left(\frac{p'}{\psi}\right)^{2}
+\frac{1}{3}\right]\theta^2
-2\left(1+\psi^2\right)\omega^2+\frac{\varrho+3p}{2} \, .
\ee
Comparing these two equations we arrive at the relation
\bea
\frac{p'}{\psi^2}\left\{\left[3\left(\frac{p'}{\psi}\right)^{2}
+\frac{1}{3}\right]\theta^2
-2\left(1+\psi^2\right)\omega^2+\frac{\varrho+3p}{2}\right\} = \hspace{40mm}
\nonumber \\
= \left[ 3\left(
\frac{p'}{\psi}\right)^{2}+\frac{1}{3}-(\varrho+p)\frac{p''}{p'}\right]
\omega^{2} , \hspace{5mm} \label{dosli}
\eea
whose time propagation provides an expression containing $p'''$.
On the other hand, the $\omega^{a}\nabla^{}_{\!a}$ and
$\omega^{ab}\nabla^{}_{\!b}$ derivatives of (\ref{dosli}) lead to two new
equations containing also $p'''$. By eliminating $p'''$ from these
three equations two different cases appear

\vspace{3mm}

\noindent $\bullet\,$ \underline{Case i)}: 
\be
2(1+\psi^2)p'+ \left[ 3\left(
\frac{p'}{\psi}\right)^{2}+\frac{1}{3}-(\varrho+p)\frac{p''}{p'}\right]
\psi^2 = 0 . \label{necua}
\ee

\vspace{3mm}

\noindent $\bullet\,$ \underline{Case ii)}:
\be
\frac{p'}{\psi}\omega^{a}\nabla^{}_{\!a}\omega^2 = 2\left(p'-\frac{2}{3}\right)
\omega^{4} \, , \hspace{16mm} \omega^{ab}\nabla^{}_{\!b}\omega^2 = 0.
\label{dosec}
\ee

\vspace{3mm}

In case i), substituting (\ref{necua}) into (\ref{dosli}) we obtain
\be
\left[3\left(\frac{p'}{\psi}\right)^{2}
+\frac{1}{3}\right]\theta^2 + \frac{\varrho+3p}{2} = 0, \label{jode}
\ee
and time--propagation of this equation leads to
\be
\theta \left(\varrho +p\right) \left(1+3 p'\right)=0 \, . \label{finale}
\ee
Here, the second factor cannot vanish because of the assumptions of the
theorem. If the third factor vanished, Eqns.\ (\ref{necua}) and (\ref{jode})
would imply $1+\psi^2 = 0$, which is impossible. Therefore, in case i) we must
have $\theta =0$, proving the theorem for this case.

In case ii) the time propagation of Eqns.\ (\ref{dosec}) does not give 
additional information. Consider instead the following Ricci identities
for the vorticity
\be
2\nabla^{}_{\![a}\nabla^{}_{\!b]}
\omega^{c} = R^{c}_{\;dab}\omega^{d} \hspace{3mm} \Longrightarrow
\hspace{3mm} \nabla^{}_{\!c}\left(\omega^{b}\nabla^{}_{\!b}\omega^{c}\right)
-\nabla^{c}\omega^{b}\nabla^{}_{\!b}\omega^{}_{c}-\omega^{b}
\nabla^{}_{\!b}\nabla^{}_{\!c}\omega^{c} = R^{}_{bd}\omega^{b}\omega^{d} ,
\label{thek}
\ee
Taking into account that the combination of (\ref{notie}) and
(\ref{dosec}) gives
\be
\omega^{b}\nabla^{}_{\!b}\omega^{a} = \omega^{2}\left[\frac{\theta}{3}
u^{a} + \left(p'-\frac{2}{3}\right)\omega^{a}\right] ,
\ee
and considering the relations obtained until now, Equation (\ref{thek}) becomes
\be
(\varrho+p)\omega^2+W^{ab}W^{}_{ab} = 0,
\hspace{5mm}
W^{}_{ab} \equiv Q^{\;c}_{(a}Q^{\;d}_{b)}\nabla^{}_{\!c}\omega^{}_{d}-
\frac{1}{2}Q^{}_{ab}Q^{cd}\nabla^{}_{\!c}\omega^{}_{d} , \label{otra}
\ee
where $Q^{}_{ab} \equiv \omega^{-2}\omega^{\;c}_{a}\omega^{}_{bc}$
is the projector orthogonal to both $\vec{u}$ and $\vec{\omega}$. 
Taking into account that 
\be
\dot{Q}^{}_{ab} = 0,
\ee
the time propagation of (\ref{otra}) leads again to the simple relation
(\ref{finale}). If $\theta =0$ we are done. Finally, if $p'=-1/3$, then the
time propagation of the relation (\ref{dosli}) leads, since
$1+\psi^2 \neq 0$, to $\theta \omega =0$, which finishes the proof.

\section{Comparison with the Newtonian case.}
\subsection*{Newtonian cosmology}
By a {\em Newtonian cosmology} is meant 
\begin{enumerate}
\item A manifold $M=\mbox{E}^3 \times \mbox{R}$, where $\mbox{E}^3$ is 
Euclidean 3-space.  For every event $({\bf x},\,t)$ in $M$ the fourth
coordinate $t$ registers its absolute time coordinate.
\item An assignment of three functions $\varrho({\bf x},t)$, 
$p({\bf x},t)$ and $\phi({\bf x},t)$ on $M$, called respectively {\em
density}, {\em pressure} and {\em gravitational potential}.
\item A differentiable 3-vector field ${\bf v}({\bf x},t)$ (the {\em
velocity vector field}), defined on each Euclidean space 
$t = \mbox{const}$, whose components depend differentiably on $t$.
\item The following equations hold relating the quantities defined above 
(summation convention adopted even though all indices are subscripts, and
units have been changed in order to restore $G$) :
\be
\dot{\varrho} \equiv \frac{d\varrho}{dt} \equiv
\frac{\partial \varrho}{\partial t} + v_i \varrho_{,i} =
-\varrho \, v_{i,i} \label{nrhodot}
\ee
\be
\dot{v}_i \equiv \frac{dv_i}{dt} \equiv
\frac{\partial v_i}{\partial t} + v_j v_{i,j} =
-\phi_{,i} -\frac{1}{\varrho}p_{,i} \label{nvidot}
\ee
\be
\phi_{,ii} = 4\pi G \varrho. \label{nphiii}
\ee
\end{enumerate}

\subsection*{Homogeneous Newtonian cosmologies}
A Newtonian cosmology will be called {\em homogeneous} \cite{SZER}
if $\varrho$ and $p$ have no spatial dependence, $\varrho=\varrho(t)$,
$p=p(t)$ and the velocity vector field depends linearly on the spatial 
coordinates,
\be v_i = V_{ij}(t) x_j \label{nVij} \ee
for some matrix of components $[V_{ij}]$ depending only on time. From
(\ref{nvidot}) it follows that $\phi_{,i}$ also is linearly dependent on
$x_i$,
\be \phi_{,i} = -f_{ij}(t)\, x_j \label{nfij}\ee
where
\be f_{ij} = \dot{V}_{ij} + V_{ik}V_{kj}\,. \label{nfijdef}\ee

Decomposing $v_{i,j}$ into its standard irreducible parts
\be v_{i,j} = \theta \frac{1}{3}\delta_{ij} + \sigma_{ij} +\omega_{ij} \label{ndecomp}\ee
where
\bea \theta & = & v_{i,i} \;=\;\mbox{expansion} \label{ntheta}\\
\sigma_{ij} & = & \frac{1}{2}(v_{i,j}+v_{j,i}) - \frac{1}{3} \theta \delta_{ij} \;=\;\mbox{shear} \label{nshear}\\
\omega_{ij}& = & \frac{1}{2}(v_{i,j}-v_{j,i}) \;=\;\mbox{rotation},
\eea
it is found in the homogeneous case that
\be \theta = V_{kk} = \theta(t),\hspace{5mm}\sigma_{ij} = V_{(ij)} - \frac{1}{3}\theta \delta_{ij} = \sigma_{ij}(t),\hspace{5mm} \omega_{ij} = V_{[ij]} = \omega_{ij}(t). \label{nhomog}\ee
Now suppose that one sets $\sigma_{ij} = 0$ in the homogeneous case, then
Eqns (\ref{nrhodot}-\ref{nphiii}) reduce to
\be \dot{\varrho} = -\theta \varrho \label{nhrhodot}\ee
\be \dot{\omega}_{ij} = -\frac{2}{3} \theta \omega_{ij} \label{nhomegadot} \ee
\be \dot{\theta} = -\frac{1}{3}\theta^2 + 2 \omega^2 -4\pi G \rho \label{nhthetadot} \ee
\be f_{ij} = \omega_{ik}\omega_{kj} + \frac{2}{3}(\omega^2 - 2 \pi G \varrho)\delta_{ij} \label{nhfij} \ee
where $\omega^2 = \frac{1}{2}\omega_{ij}\omega_{ij}$.

Define a variable $R=R(t)$ such that $\theta = 3\dot{R}/R$, then it is a 
straightforward matter to integrate (\ref{nhrhodot}) and (\ref{nhomegadot}) 
\be \varrho=\varrho(t_0) \,R^{-3}, \hspace{5mm} \omega_{ij} = \omega_{ij}(t_0)\,R^{-2}\ee
where $R(t_0) = 1$ and substitution in (\ref{nhthetadot}) gives
\be \ddot{R} = \frac{-4\pi G}{3} \frac{\varrho(t_0)}{R^2} +\frac{2\omega^2(t_0)}{3R^3}.\ee
This integrates to give the Heckmann-Sch\"{u}cking equation \cite{HECS}
\be \dot{R}^2 = \frac{8\pi G\varrho(t_0)}{3R} - \frac{2\omega^2(t_0)}{3R^2} + C \label{heck-schuck}\ee
where $C$ is an arbitrary constant.  Solutions of this equation represent
shear-free Newtonian cosmologies which are in general both expanding 
($\dot{R} \neq 0$) and rotating ($\omega^2(t_0) \neq 0$).  The 
gravitational potential is simply read off from (\ref{nhfij}) using 
(\ref{nfij}).  A point of interest is that whenever $\omega^2(t_0) \neq 0$
there is no singularity since $R(t)$ always has a minimum value $R_0 >0$ 
where $\dot{R} = 0$.  Thus the big-bang singularity of the Newtonian 
equivalent to the standard FLRW models of general relativity (namely, the 
case $\omega^2=0$), is easily avoided by giving the model an arbitrarily
small amount of rotation.  This is quite contrary to the case in general 
relativity where singularities are generic to all spatially homogeneous
models \cite{HAE2}.

\subsection*{Newtonian version of our proof}
These homogeneous shear-free solutions are independent of the pressure, which
may as well be set equal to zero (or constant).  The question naturally arises 
as to why there are shear-free dust solutions in Newtonian theory having
$\theta\omega \neq 0$, when Ellis's theorem ensures that none exist in general
relativity.  This difference in the two theories is both surprising and
interesting, since Ellis's is a purely local result and is completely
independent of the strength of the gravitational field.   One would then expect
it to hold in the weak field Newtonian limit, yet clearly it does not.  What
then is going wrong in this limit?

To investigate this question we propose to follow the proof of Ellis's
theorem given here in Section 2, but try to apply it as closely as possible 
to Newtonian cosmology in an attempt to pinpoint exactly where it is that 
the proof fails.

Consider a general pressure-free ($p=0$) Newtonian cosmology (not
necessarily homogeneous) having $\sigma_{ij}=0$, i.e.
\be v_{i,j} = \frac{1}{3}\theta \,\delta_{ij} + \omega_{ij} \label{n2cod2} \ee
\be \dot{v}_i = -\phi_{,i} \label{nnew} \ee
where the Newtonian time evolution operator is given by
\[\dot{} = \frac{d}{dt} = \frac{\partial}{\partial t} + v_i\, \frac{\partial}{\partial_i}\]
and
\be \phi_{,ii} = \frac{1}{2}\varrho \label{nphiii2}.\ee
In this last equation, (\ref{nphiii}) has been recast in units such that
$8\pi G =1$ which has been done in order to bring the Newtonian and
general relativistic equations into closer comparison.  Eqns. 
(\ref{n2cod2}) and (\ref{nnew}) correspond to the earlier equations
(\ref{2cod2}) and (\ref{new}), while Eqn.\ (\ref{nrhodot}) reads
\be \dot{\varrho} + \varrho \theta =0 \label{nrhodot2},\ee
which is identical with the relativistic equation (\ref{2ropu}) when $p=0$.

A useful identity is
\be \frac{d}{dt}\frac{\partial}{\partial x_i} = \frac{\partial}{\partial x_i}\frac{d}{dt} -\frac{1}{3}\theta \frac{\partial}{\partial x_i} + \omega_{ij} \frac{\partial}{\partial x_j} \label{ndotid}\ee
which takes the place of (\ref{dotid}). Now perform the time evolution
of Equation (\ref{n2cod2}) and use (\ref{ndotid}) to arrive at
\be \dot{\omega}_{ij} = -\frac{2}{3}\theta \omega_{ij}, \label{n2ropo}\ee
\be \dot{\theta} = -\frac{1}{2}\varrho -\frac{1}{3}\theta^2 + 2\omega^2, \label{n2rayc} \ee
and
\be E_{ij} \equiv \phi_{,ij}-\frac{1}{3}\phi_{,kk}\delta_{ij} = \omega_{ik}\omega_{jk} -\frac{2}{3}\omega^2 \delta_{ij} = -\omega_i \omega_j  + \frac{1}{3}\omega^2 \delta_{ij}, \label{n3elwe}\ee
where $\omega_i = \frac{1}{2}\epsilon_{ijk}\omega_{jk}$.
These equations are clearly Newtonian versions of the relativistic 
equations (\ref{2ropo}), (\ref{2rayc}) and (\ref{3elwe}) when $\acc^b=0$.
>From (\ref{n2ropo}) there follows the equivalent of (\ref{wdot}), i.e.
\be \dot{\omega}= -\frac{2}{3}\theta\,\omega. \label{nwdot}\ee
Applying $v_{i,jk} = v_{i,kj}$ to (\ref{n2cod2}) gives the cyclic identity
\be \omega_{ij,k}+\omega_{jk,i}+\omega_{ki,j}=0, \label{ncyclic}\ee
and
\be \omega_{jk,i} = \frac{1}{3}(\theta_{,k} \delta_{ij} - \theta_{,j}\delta_{ik}) \label{nwjki}.\ee
Contracting Eq.\ (\ref{ncyclic}) with $\omega_{ij}$ gives
\be \omega \omega_{,k} = \omega_{ij}\omega_{ik,j} \label{nwwj} \ee
while contraction of (\ref{nwjki}) over $ij$ gives the Newtonian equivalent 
of the important equation (\ref{3main}),
\be \theta_{,k} = \frac{3}{2} \omega_{ik,i}. \label{n3main}\ee
Finally contract Eq.\ (\ref{nwjki}) with $\omega_{ij}$ and use (\ref{nwwj}) 
to give
\be \omega_{ij}\theta_{,j} = 3\omega\,\omega_{,i}.\label{nomtheta}\ee
Now perform the time evolution of (\ref{n3main}), apply (\ref{ndotid})
to both sides and use the identities (\ref{n2ropo}), (\ref{n2rayc}), (\ref{nwwj}), (\ref{n3main}) and (\ref{nomtheta}), results in the key equation
\be \varrho_{,k} - 8 \omega\,\omega_{,k} +\omega_{ki}\theta_{,i} =0 \label{nkey}\ee
which is basically the same as Equation (\ref{key}).  Actually the 
Newtonian case is rather stronger than that in general relativity since 
the further identity (\ref{nomtheta}), which has no relativistic equivalent, 
results in
\be \varrho_{,k} = 5\omega\,\omega_{,k} = \frac{5}{3} \omega_{ki}\theta_{,i}. \label{nextra}\ee
At first sight this is a rather surprising result, since Ellis's theorem
is a theorem concerning restrictions on solutions of general 
relativity, so one might expect that Newtonian theory should place {\em 
less} stringent restrictions on its solutions. 

In any case one can continue in exactly the same way as the proof given
for Theorem 1 to arrive at a Newtonian version of Equation (\ref{F}), and its contracted version
\be
\left[\frac{32}{9}\theta^2\omega^2+ \left(\varrho -\frac{16}{3}\omega^2
\right)\left(\frac{\varrho}{2}-2\omega^2\right)\right]
\omega_i \theta_{,i} = 0 . \label{nF2}
\ee

\subsection*{Newtonian equivalent of Lemmas 1 and 3}
In the final stages of the proof of Theorem 1 use is made of Lemmas 1 and
3.  The Newtonian equivalent of Lemma 3, with $p=0$ is\\[2mm]
{\bf Lemma $\bf 3'$}: \hspace{2mm} {\em If $\varrho =k\,\omega^2$ for a 
non-zero constant $k$, then $\theta \omega=0$.}\\[2mm]
This lemma holds in Newtonian cosmology, for on using (\ref{nrhodot2})
and (\ref{nwdot})
\[\dot{\varrho} = -\theta \varrho = 2k\,\omega\dot{\omega} = -\frac{4}{3}k\,\theta\omega^2.\]
Hence \[\frac{1}{3}k\,\theta \omega^2=0\]
which proves the result.

Lemma 1 however is a different matter. The obvious Newtonian translation
would be the following:\\[2mm]
{\em If a function $f({\bf x},t)$ has zero gradient, $f_{,i} = 0$ then $f$ is
constant or $\omega_{ij} = 0$.}\\[2mm]
Of course a function whose gradient vanishes may still be dependent on
time, $f=f(t)$.  This can have no bearing whatsoever on the rotation of
the velocity vector field.  Hence as it stands, this statement cannot
be true.  It is worth trying to understand why the limiting process does
not apply for this lemma.

Suppose we have a metric $g_{ab}=\eta_{ab} + h_{ab}$ having energy-stress
tensor $T_{ab}$ given by Equation (\ref{2temd}) where (restoring $c$)
\be u_a=\left(\frac{v_i}{c}, -1 + O\left(\frac{v^2}{c^2}\right)\right), \hspace{3mm} |h_{ab}| = O\left(\frac{v^2}{c^2}\right), \ee
where $v^2 = v_iv_i$.  The Newtonian limit is the limit of all equations
found on letting $c\,\rightarrow\,\infty$.  For example, the rotation is 
given by
\be \omega_{ij} = \lim_{c\rightarrow \infty} c\, u_{[i,j]} = v_{[i,j]}, \label{omlim}\ee
as to be expected.  However the equation $P_{a}^{\;b}\nabla_b f = 0$ 
reduces in its first three components to
\be f_{,i} = -\frac{1}{c^2}\dot{f} v_i \label{fi}\ee
where $\displaystyle \dot{f} = \frac{\partial f}{\partial t} +f_{,i}v_i$.
The limit as $c\rightarrow \infty$ of this equation is clearly $f_{,i}=0$,
from which no conclusions can be made concerning $\omega_{ij} =v_{[i,j]}$. 
In general relativity a conclusion can be reached because Equation 
(\ref{ugrad}) couples $u_a$ and $\nabla_a f$, but in taking the Newtonian 
limit $v_i$ and $f_{,i}$ become decoupled.

Physically one understands it like this. If $\omega_{ab}=0$ then the 
4-velocity field $u_a$ is hypersurface orthogonal, and in a sense there is 
a universal time coordinate $t$ defined such that $u_a \propto \nabla_a t$
(in the lemma, $f$ could act as such a time coordinate). However in 
Newtonian theory such a universal time coordinate is {\em always} defined and 
the condition $\omega_{ij}=0$ is equivalent to $v_i = \phi_{,i}$ for some 
scalar field $\phi$ having in general nothing whatsoever to do with 
the universal time. For example, $\phi$ may just depend on the 
spatial coordinates $x,y,z$. Lemma 1 is truly a relativistic result, 
having no Newtonian equivalent whatsoever.

\subsection*{Final stages of the proof}
Returning to the proof of Theorem 1, we attempt to follow the
argument given in the final stages. Consider the two cases arising from
Equation (\ref{nF2}).  

Case $1'$): $\omega_i \theta_{,i} \neq 0$.  In place of Equation 
(\ref{cond1a}) the corresponding Newtonian argument arrives at
\be \left(\frac{32}{9}\omega^2 -\varrho\right)\omega^2 =0. \ee
Lemma $3'$ immediately applies, showing that $\theta \omega=0$

Case $2'$) $\omega_i \theta_{,i} = 0$. The argument can again be continued 
as in Theorem 1.  The Newtonian version of (\ref{F3}) is
\be \frac{\theta}{2}(29\omega^2 - 6\varrho)\omega_{ij}\theta_{,j} = 
\left[\frac{58}{3}\theta^2\omega^2 + \left(2\omega^2 - \frac{\varrho}{2} \right)\left(29\omega^2-6\varrho \right)\right]\theta_{,i}. \label{nF3}\ee
Now clearly $\omega_{ij}\theta_{,j}$ is orthogonal to $\theta_{,i}$ (since
$\omega_{ij}\theta_{,i}\theta_{,j}=0$ and $\omega_{ij}=-\omega_{ji}$),
whence the left hand side of (\ref{nF3}) vanishes.  If $\theta=0$ the
proof is over, and by Lemma $3'$ this is also true if $29\omega^2 - 6\varrho 
=0$. If $\omega_{ij}\theta_{,j} =0$ then $\epsilon_{ijk} \omega_j 
\theta_{,k} =0$, and combined with $\omega_i\theta_{,i} = 0$ it follows that
$\omega_i =0$ or $\theta_{,i} =0$.  The former implies $\omega=0$ and we 
are done, while the latter implies by (\ref{nextra}) that
\be \varrho_{,i}=\omega_{,i} = 0. \label{n0}\ee
However it is not possible to proceed any further.  It is at this very last 
step that our attempt at a proof comes to an end 
since there is no equivalent to Lemma 1. The limiting process has broken 
down for the reasons outlined above.

\subsection*{All Newtonian shear-free universes}
To conclude this Newtonian discussion, we propose to show that all
pressure-free, shear-free Newtonian universes are in fact homogeneous.
If $\sigma=0$ then the above discussion has shown that $\theta
\omega =0$ or $\theta_{,i}=0$. In either case we see from Equation 
(\ref{nextra}) that
\be \varrho_{,i} = \theta_{,i} = \omega_{,i} = 0, \label{nall0}\ee
whence
\be \varrho=\varrho(t), \hspace{3mm} \theta = \theta(t), \hspace{3mm} \omega=\omega(t). \label{nallt}\ee
>From (\ref{n2cod2}), using (\ref{nall0}) and $v_{i,jk}=v_{i,kj}$ we see
that $\omega_{ij,k}=\omega_{ik,j}$, and substituting in Equation 
(\ref{ncyclic}) gives $\omega_{jk,i} = 0$, i.e.\ $\omega_{jk} = 
\omega_{jk}(t)$.  Hence
\[ v_{i,j} = \frac{1}{3}\theta(t)\delta_{ij} + \omega_{ij}(t) = v_{i,j}(t)\]
from which it follows that 
\[v_i = V_{ij}x_j \hspace{5mm}\mbox{where } V_{ij}=\frac{1}{3}\theta(t)\delta_{ij} + \omega_{ij}(t) = V_{ij}(t).\]
All conditions for a homogeneous cosmology are now satisfied, and it 
follows that all shear-free Newtonian cosmologies are homogeneous (and are
therefore in the Heckmann--Sch\"{u}cking class).  What this means is
that although Ellis's theorem does not hold in Newtonian cosmology,
it ``nearly'' holds in the sense that the only solutions which 
violate the theorem are homogeneous.

\section{Conclusions}
Ellis's theorem and its various extensions have up till now always required 
the use of tetrads in their proof.  In this paper we have shown that it is 
possible to prove such results by methods which are totally covariant in 
character, and which do not hinge on setting up specific tetrads or 
coordinate systems.  While in this paper we do not go further than showing 
already known results, our proofs of these theorems have considerable
elegance and lend themselves readily for extension to more difficult cases.

Another advantage of our proofs has been the light they shed on the puzzle  
of the Newtonian equivalence.  It seems that the basic reason these results  
do not hold in Newtonian cosmology is that there is already a well-defined  
universal time in that theory. In relativistic cosmologies, there will only  
be a universal time, defined for example by the energy-stress tensor, if the 
rotation of the 4-velocity field vanishes.  Thus a rotating universe can  have
no non-constant scalar field whose spatial gradient with respect to the
4-velocity vanishes (the content  of Lemma 1). However in the Newtonian limit
this statement is quite untrue,  since there exist functions of the universal
time alone, even when the  3-velocity is rotating.

\end{document}